\documentclass[twocolumn]{article}
\usepackage{cite}
\usepackage{amsmath,amssymb,amsfonts}
\usepackage{algorithmic}
\usepackage{graphicx}
\usepackage{textcomp}
\usepackage{times}
\usepackage{caption}
\usepackage{booktabs} 
\usepackage{float}
\usepackage{url}
\usepackage{enumitem}
\usepackage[table]{xcolor}

\usepackage{dcolumn}
\newcolumntype{d}[1]{>{\centering\arraybackslash}D{.}{.}{#1}}
\usepackage{siunitx}

\definecolor{myLightGray}{gray}{0.85}
\definecolor{myWhite}{gray}{1.0}

\def\BibTeX{{\rm B\kern-.05em{\sc i\kern-.025em b}\kern-.08em
    T\kern-.1667em\lower.7ex\hbox{E}\kern-.125emX}}

\title{A Unified Architecture for N-Dimensional Visualization and Simulation:
4D Implementation and Evaluation including Boolean Operations}
\author{Hirohito Arai \\ Ritsumeikan Uji High School}
\date{}
    
\begin{document}
\maketitle
\begin{abstract}
This paper proposes a unified software architecture for visualization and simulation based on a design targeting an $N$-dimensional space.
The contributions of this study are twofold. First, it presents an architectural configuration that integrates multiple processes into a single software architecture: Quickhull-based convex hull mesh generation, Boolean operations, coordinate transformations for high-dimensional exploration (pose transformation and view transformation), and hyperplane slicing for visualization. Second, it defines “Plex” (\texttt{.plex}) as a file format intended for the exchange of $N$-dimensional mesh data.
The proposed approach adopts an approximate implementation that tolerates numerical errors and prioritizes implementation transparency over guarantees of numerical rigor.
The experimental results and evaluations presented in this paper are limited to a 4D implementation; no evaluation is conducted for $N>4$, and the discussion is restricted to stating that the architecture itself has a dimension-independent structure.
This paper also proposes an interaction design for high-dimensional exploration based on FPS navigation.
As an input example involving shape changes over time, a non-rigid body simulation based on XPBD (Extended Position Based Dynamics) is integrated into the 4D implementation.
Experimental results confirm that the 4D implementation runs on a single PC.
\end{abstract}

\section{Introduction}\label{sec:introduction}
This paper builds on prior work on visualization and simulation in high-dimensional spaces. However, examples remain limited that integrate Quickhull-based convex hull mesh generation, Boolean operations, coordinate transformations for high-dimensional exploration (orientation and view transformations), and hyperplane slicing for visualization into a single software architecture, under the assumption of an iterative workflow of user actions and result inspection.

This study proposes a unified architecture intended for use by researchers and developers. When exchanging high-dimensional mesh data across projects, the availability of common formats is limited. Consequently, project-specific formats or extensions of existing formats are often employed.

This study has two main contributions. First, it presents a configuration that integrates the above processing components into a single software architecture and evaluates a four-dimensional implementation through experiments. Second, it defines “Plex” (\texttt{.plex}) as a file format intended for the exchange of $N$-dimensional mesh data.

The architecture is designed for $N$-dimensional space; however, the implementation and evaluation in this study are limited to 4D. In this paper, interactivity is defined as a workflow that presupposes repeated cycles of user actions and result inspection, in contrast to offline batch processing in which input conditions are fixed in advance and executed in a single run. Hereafter, the term “interactive,” including its adjectival and adverbial usages, refers to usage scenarios that satisfy or require this notion of interactivity.

\section{Related Work}
Prior work on visualization and simulation in high-dimensional spaces is often reported in terms of individual elemental techniques.

For convex hull computation applicable to high dimensions, the pivoting method \cite{pivoting} and Quickhull \cite{quickhull} are known as representative approaches. In implementations that ensure robustness against inputs with numerical errors or degenerate configurations, these algorithms can incur increased computational cost due to exact predicates and additional handling.

Convex hull computation is part of the fundamental problem set in high-dimensional computational geometry, and textbooks systematically organize problems such as convex hulls, halfspace intersection, and arrangements \cite{Computationalgeometry}.

As existing methods for high-dimensional visualization, there are approaches based on projection to lower-dimensional spaces. Cook et al. provide an overview of methods and design issues for exploring and presenting high-dimensional information via 2D projections \cite{Cook2008}. Dimensional stacking has also been surveyed as a class of visualization techniques that assigns each dimension to a nested 2D layout \cite{keim2002information}.

These works mainly survey visualization techniques. In contrast, on the implementation side, architectures and frameworks for high-dimensional visualization have also been proposed. For mesh-based approaches, a GPU-based 4D visualization architecture has been proposed \cite{GL4D} and discussed from the viewpoint of rendering and visualization of 4D shapes. As a software framework for computations on convex polyhedra, “polymake” has been reported \cite{polymake}. It provides visualization, exemplified by Schlegel diagrams, together with the underlying algorithmic processing. However, these prior examples differ in scope from the unified environment targeted in this paper, which aims to integrate not only visualization but also other processing in an interactive manner.

For Boolean operations on 3D meshes, methods based on intersection resolution and spatial partitioning have been reported; Zhou et al. present a unified framework that handles union, intersection, and difference based on cell decomposition and extraction using winding number vectors \cite{Mesharrangements}. In the 3D setting, implementation examples of interactive Boolean operations have also been reported \cite{RobustMeshBooleans}. CGAL adopts a design policy that emphasizes geometric robustness \cite{CGAL}. Under this policy, CGAL and related libraries provide set operations in 2D and 3D, exemplified by Nef polyhedra, whereas implementation reports of Boolean operations on explicit boundary meshes in 4D and higher dimensions are limited.

For set operations in 4D and higher dimensions, libraries for high-dimensional computation based on polyhedral representations have been proposed \cite{polyhedrallibrary}. However, these assume representations based on halfspace constraints or cell decompositions, rather than explicit boundary meshes. They also typically assume offline computation with exact arithmetic, and implementation reports aimed at interactive operating environments remain limited.

In physical simulation, non-rigid body simulation based on XPBD has been reported in 3D space \cite{XPBD}, and there are also studies that handle rigid bodies in 4D and higher dimensions \cite{nrigid}. However, reports whose primary goal is to visualize non-rigid objects in 4D and higher dimensions and handle them interactively remain limited.

\section{Proposed Method}
\subsection{Overview of the Overall Architecture}\label{overallabstract}
\begin{figure}[ht]
    \centering
    \includegraphics[width=\columnwidth]{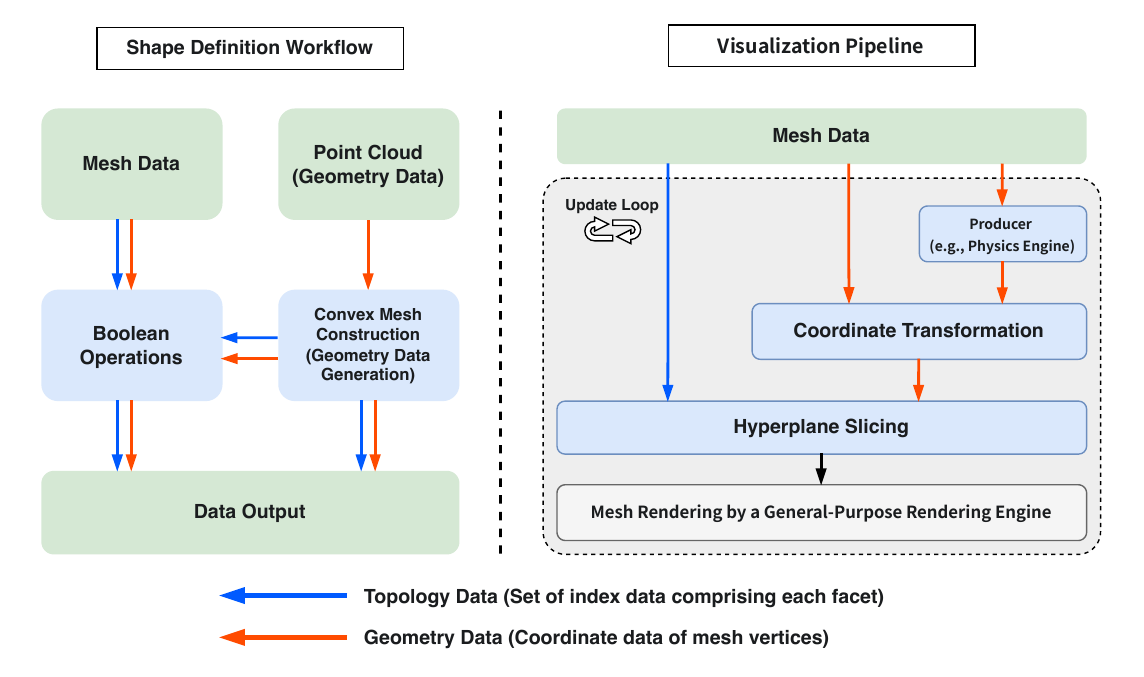}
    \caption{An overview of the proposed architecture. The left side shows the static shape definition workflow, and the right side shows the dynamic visualization pipeline.}
    \label{fig:FrameworkFlow}
\end{figure}
This section describes the technical details of the proposed architecture. We first provide an overview of the overall architecture, and then describe in detail the algorithms that constitute each function and their implementations.

As shown in Fig. \ref{fig:FrameworkFlow}, the overall structure and data flow of the architecture consist of multiple components: convex hull mesh generation, Boolean operations, coordinate transformations, and hyperplane slicing.
The convex hull mesh generation component employs a Quickhull-based algorithm to construct a list of facet indices that define the topology of an $N$-dimensional object from the input point set data.
The Boolean operation component dynamically modifies the topology and geometry of the mesh based on user interactions.
The coordinate transformation component converts vertex coordinates into a camera-referenced coordinate system by applying a view transformation based on the viewpoint, after incorporating orientation transformations updated through user operations.
The hyperplane slicing component applies hyperplane slicing to the output mesh from the coordinate transformation component and generates and renders 3D cross-sections within the update loop.

In this study, geometric algebra is adopted as the mathematical foundation for uniformly describing geometric entities such as $N$-dimensional vectors, planes, and rotations. Introductory explanations of geometric algebra and its use in computer science and computer graphics can be found in \cite{vince2008}, systematic treatments in \cite{doran2003}, and survey-style overviews in \cite{macdonald2017}.

\subsection{Design Philosophy and Architectural Overview}\label{designPhilosophy}
The design policy of this architecture emphasizes three principles: algorithmic transparency, minimization of dependence on external libraries, and separation of topology and geometry. This policy is based on a design decision that prioritizes simplification of description and implementation over performance optimization, so that researchers and developers can trace the structure of the algorithms.

Based on this design policy, all computational geometry logic is unified under a CPU-based implementation written in standard C\#. While GPU-based parallelization can deliver high performance for specific tasks, it introduces issues of code complexity and hardware dependency. The CPU-based C\# implementation is intended to enhance platform independence and algorithmic transparency.
In addition, this paper adopts an approximate implementation that tolerates numerical errors and does not guarantee numerical robustness arising from floating-point arithmetic, which is positioned as a subject for future work.
The 4D implementation in this study was developed in C\# on the Unity engine and is released as open-source software \cite{hirohito_arai_2026_18247896}.

In this architecture, a design is adopted in which the vertex index sequences that define facets (topology) and the vertex coordinate arrays (geometry) are managed independently. This design assumes scenarios in which time-varying geometry is provided as input for a fixed topology. Details are described in Section~\ref{TopologyandGeometry}.

\subsection{N-Dimensional Convex Hull Mesh Generation}
\begin{figure*}[t]
    \centering
    \includegraphics[width=\textwidth]{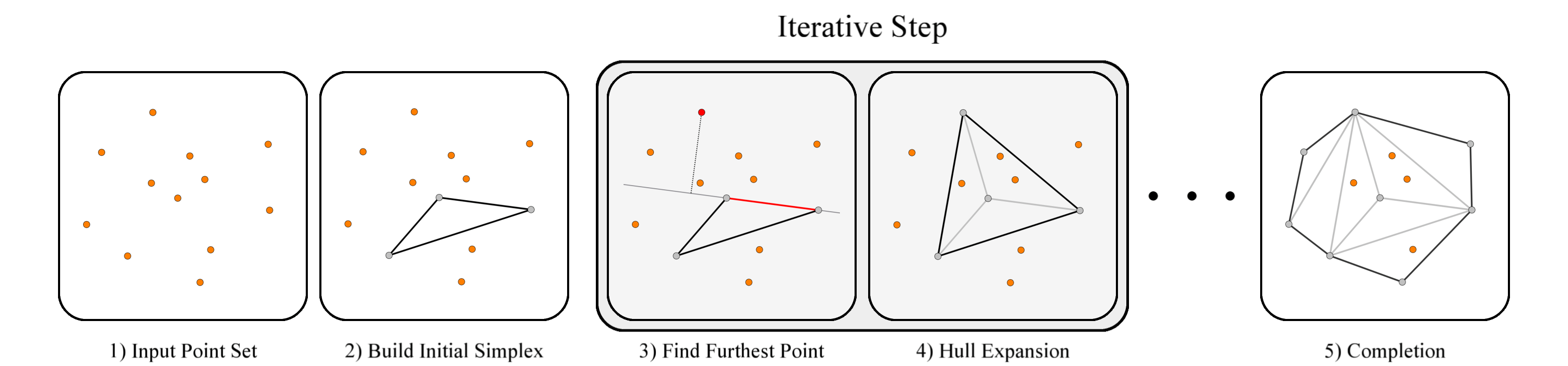}
    \caption{Stages of vertex management and convex hull construction in Direct Quickhull. Orange vertices are unprocessed, gray vertices are processed, gray facets are interior, and black facets form the outer hull. In the iterative step, the convex hull is expanded using a selected facet from the hull (red) and its farthest point (red).}
    \label{fig:quickhull}
\end{figure*}

\subsubsection{Comparison with Existing Methods}\label{Diff of methods}
This component is centered on a self-contained C\# implementation of the classical $N$-dimensional Quickhull algorithm \cite{quickhull}. A standard implementation would perform an optimization called candidate point pruning to maximize computational efficiency. In this strategy, each facet maintains its own list of candidate points, and points that have been absorbed into the interior of the convex hull are excluded from further computation; this is effective for volume-like point sets containing many interior points. However, in this architecture, the primary use cases are basic convex hull mesh construction for initial shapes and dynamic simplex tessellation (described in Section~\ref{simplex_filling}), and it often handles small surface-like point sets with only a few dozen vertices and few interior points. In this context, the performance gains from pruning are limited, whereas increased implementation complexity entails a risk of reduced code maintainability.

Therefore, this architecture deliberately removes the pruning mechanism in light of this trade-off. As shown in Fig.~\ref{fig:quickhull}, it linearly searches a single global set of candidate points to find the farthest point, and removes from the candidate set only those points that are used as convex hull vertices. In this paper, we refer to this approach as “Direct Quickhull.”

\subsubsection{Calculating N-Dimensional Normals and Visible Facets}
In Direct Quickhull, facet normal vectors are computed iteratively, and this computation can become a dominant factor in the overall processing time. One possible approach for computing normals is to use null-space computation; however, this typically requires singular value decomposition (SVD) or similar methods, which can impose a substantial computational burden in the iterative calculations assumed for interactive use.
To address this issue, this architecture introduces the concepts of the wedge product and the dual from geometric Algebra. First, in an $N$-dimensional space $\mathbb{R}^N$, we define a facet from its $N$ vertices $\{\mathbf{p}_0, \mathbf{p}_1, \dots, \mathbf{p}_{N-1}\}$ and construct the $(N-1)$ linearly independent vectors that span the facet.

\begin{equation} \label{eq:spanning_vectors}
\mathbf{v}_i = \mathbf{p}_i - \mathbf{p}_0 \quad (\text{for } i = 1, \dots, N-1)
\end{equation}

Next, we calculate the wedge product of these vectors.

\begin{equation} \label{eq:wedge_product}
\mathbf{B} = \mathbf{v}_1 \wedge \mathbf{v}_2 \wedge \dots \wedge \mathbf{v}_{N-1}
\end{equation}

In geometric Algebra, this $(N-1)$-blade $\mathbf{B}$ is a mathematical entity that represents the oriented hyperplane spanned by the vectors. The operation of taking the Hodge dual of this blade, $\mathbf{B}^*$, is equivalent to finding the orthogonal complement of that hyperplane. That is, for an $(N-1)$-dimensional hyperplane, its orthogonal complement is a 1-dimensional line, and its direction vector becomes the normal vector $\mathbf{n}$ to the hyperplane.

\begin{equation}
\mathbf{n} = \mathbf{B}^*
\end{equation}

This series of operations can ultimately be implemented efficiently as the calculation of the determinant of an $N \times N$ matrix, as follows:
\begin{equation}
\mathbf{n} := \det
\begin{vmatrix}
  \mathbf{e}_1 & \mathbf{e}_2 & \dots  & \mathbf{e}_N \\
  v_{1,1}      & v_{1,2}      & \dots  & v_{1,N} \\
  v_{2,1}      & v_{2,2}      & \dots  & v_{2,N} \\
  \vdots       & \vdots       & \ddots & \vdots \\
  v_{N-1,1}    & v_{N-1,2}    & \dots  & v_{N-1,N}
\end{vmatrix}
\end{equation}

Next, it is necessary to correctly determine the orientation of the facet's normal vector. In the 3D case, the normal vectors of facets constituting the surface of a convex hull were obtained with correct orientation by computing the ordering of vertices, exploiting the right-handed property of the cross product of 3D vectors. However, the cross product is unique to 3D space. In higher dimensions, there are no geometric concepts analogous to the winding order that intuitively define a face’s orientation.

Therefore, in this architecture, this issue is addressed by exploiting the geometric property that facet normal vectors of a convex hull always point outward. The orientation is determined by computing the dot product between the facet normal vector and a vector directed from a point on the facet toward the centroid of the initial simplex constructed at the beginning of the algorithm. If the sign of the dot product is negative, the normal vector is judged to be outward-facing; if the sign is positive, the normal vector is flipped to obtain the correct outward orientation.

Next, we use the facet normal vectors to determine the visible facets, which is a necessary step in the process of expanding the convex hull.

\begin{figure}[ht]
    \centering
    \includegraphics[width=\columnwidth]{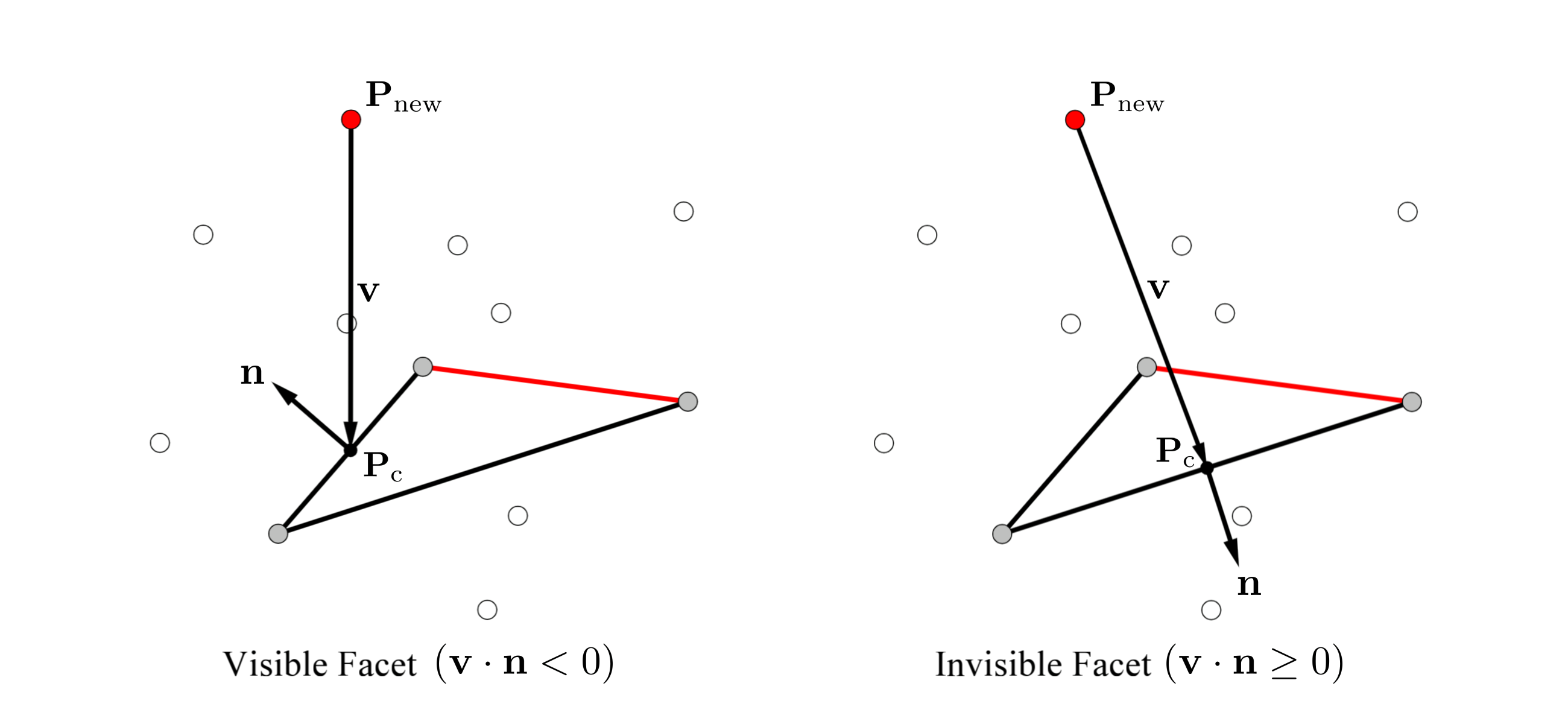}
    \caption{The logic for determining visible facets in Quickhull. This illustrates the method for identifying facets that are visible from the farthest point $\mathbf{P}_{\text{new}}$. The centroid of the facet being tested is denoted as $\mathbf{P}_{\text{c}}$, and its normal vector is $\mathbf{n}$. The vector pointing from $\mathbf{P}_{\text{new}}$ to $\mathbf{P}_{\text{c}}$ is defined as $\mathbf{v}$.}
    \label{fig:visibleFacet}
\end{figure}

As shown in Figure~\ref{fig:visibleFacet}, the facets visible from $\mathbf{P_\text{new}}$ can be found by examining the sign of the dot product of $\mathbf{v}$ and $\mathbf{n}$. The convex hull can then be expanded by connecting the new vertex $\mathbf{P_\text{new}}$ to the boundary of the identified set of visible facets (i.e., the set of ridges that are not shared among the visible facets).

\subsubsection{Simplex Filling in Convex Hull Construction}\label{simplex_filling}
In this architecture, Direct Quickhull is used for simplicial decomposition of the convex hull. In the incremental updates of Direct Quickhull, the newly added region is represented as a set of $N$-dimensional simplices generated from the new vertex $\mathbf{P}_{\text{new}}$ and the visible facets. Exploiting this property, the architecture appends the simplices generated at each update step to a simplex list and retains them, thereby obtaining the boundary facet set and the simplex list upon completion of convex hull construction.

This method is used for two purposes. First, it is used for the simplicial decomposition of a convex polytope produced by hyperplane clipping of facets in Boolean operations. Second, it is intended for determining the cross-sectional structure of facets arising in hyperplane slicing in spaces of five or more dimensions; details are described in Section~\ref{devidecuttedplane}.

\subsection{N-Dimensional Boolean Operations}
\subsubsection{Broad Phase: Candidate Pruning via Spatial Partitioning}
In $N$-dimensional Boolean operations, let the numbers of facets in the two meshes be $N_\text{A}$ and $N_\text{B}$, respectively. A brute-force intersection test over all facet pairs then has a computational complexity of $O(N_\text{A}N_\text{B})$. To avoid brute-force intersection testing, this architecture adopts a spatial partitioning grid as a broad-phase stage and extracts facet pairs that are potential intersection candidates. We compute the axis-aligned bounding box (AABB) for each facet and register the facet with the grid cells that its AABB occupies. In the subsequent narrow-phase stage, intersection tests are performed on facet pairs that share the same cell. Hierarchical methods such as BVHs and octrees could also be considered as alternatives; however, this paper prioritizes implementation simplicity and adopts a uniform grid approach.

\begin{figure*}[t]
    \centering
    \includegraphics[width=\textwidth]{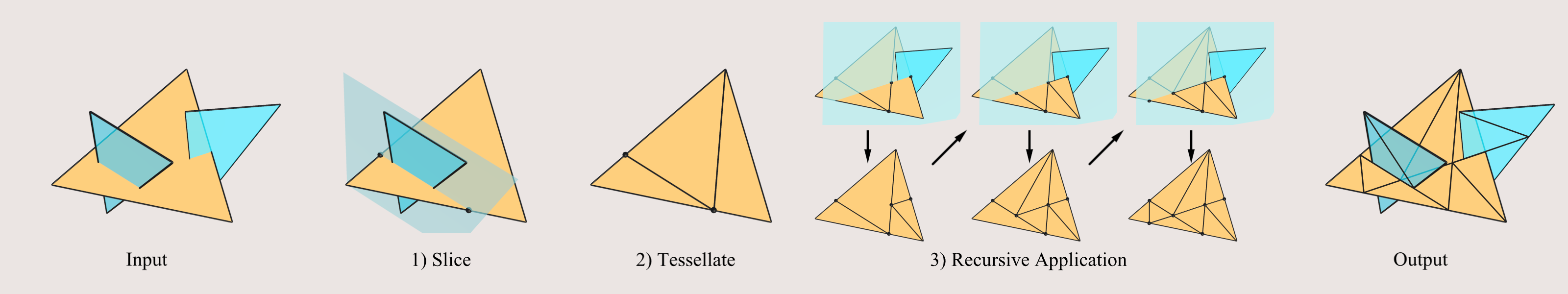}
    \caption{The sequential tessellation process of a facet in a Boolean operation (analogized in 3D). After the input facet (yellow) is clipped and tessellated by a facet from the opposing object (blue, left), each resulting facet is further tessellated by a subsequent blue facet, as shown in the center. By repeating this process for all intersecting facets, a mesh (Output) that is accurately partitioned at the boundary with the opposing object is ultimately generated.}
    \label{fig:tesselate}
\end{figure*}

\subsubsection{Narrow Phase: Facet-Facet Intersection Test}\label{IntersectionDetection}
In the narrow phase, we perform precise geometric intersection tests on the candidate pairs pruned during the broad phase. The principle of this algorithm is a generalization to $N$-dimensions of the method proposed by Tomas Möller et al. for fast triangle-triangle intersection tests in 3D \cite{Moller1997}. Its essence lies in decomposing the intersection test between two facets, $\mathcal{F}_{\text{A}}$ and $\mathcal{F}_{\text{B}}$, into two symmetric sub-problems. First, we compute the geometric intersection region, $\mathcal{I}_{\text{B}}$, between the $(N-1)$-dimensional hyperplane spanned by $\mathcal{F}_{\text{A}}$ and the facet $\mathcal{F}_{\text{B}}$. Second, we perform the same operation with the roles swapped, computing the intersection region $\mathcal{I}_{\text{A}}$ between the hyperplane spanned by $\mathcal{I}_{\text{B}}$ and the facet $\mathcal{F}_{\text{A}}$. Both of these intersection regions are $(N-2)$-dimensional convex polytopes. Finally, we determine if an intersection between the facets exists by testing whether $\mathcal{I}_{\text{A}}$ and $\mathcal{I}_{\text{B}}$ overlap within their common $(N-2)$-dimensional space.

If an intersection is found to exist, this information is utilized in the subsequent tessellation stage. Specifically, we register this intersection pair in a dictionary where one facet is the key and the other is the value. This registration is performed for all intersecting pairs, and the cutting relationships are represented as a dictionary, including cases in which a single facet has multiple intersection counterparts.

In the 4D implementation, the Separating Axis Theorem (SAT) is adopted to determine overlap between intersection regions. However, in dimensions of five or higher, the number of vertices of the $(N-2)$-dimensional polytopes representing intersection regions can increase, which in turn increases the number of candidate separating axes enumerated by SAT and may lead to higher computational cost. To address this issue, for applications in five or more dimensions, an alternative candidate is intersection testing based on the GJK algorithm.

\subsubsection{Dynamic Tessellation via Direct Quickhull}
A facet that is found to intersect in the narrow phase is clipped by the $(N-1)$-dimensional hyperplane spanned by the opposing facet, and the clipped facet is then tessellated into multiple smaller facets. As illustrated in Fig.~\ref{fig:tesselate}, a facet that is originally an $(N-1)$-dimensional simplex may, through cutting, be decomposed into an $(N-1)$-dimensional convex polytope whose number of vertices exceeds $N$. In such cases, the polytope after cutting must be subdivided into multiple smaller facets.

The core of this tessellation algorithm is to classify the vertices of the facet into the positive side (P-side) and the negative side (N-side) based on their position relative to the clipping hyperplane. The tessellation process is executed only when the facet has vertices in both half-spaces; otherwise, no tessellation is performed.

After classifying the vertices of a facet, the Direct Quickhull simplex-filling procedure is applied to a point set that combines the intersection points between the facet and the cutting hyperplane with the vertices on the positive (P) side, thereby re-subdividing the P-side region into multiple smaller facets. The same operation is then applied to the intersection points together with the vertices on the negative (N) side, and the N-side region is re-subdivided. Through these two processes, the facet is re-subdivided so as to be consistent with the cutting plane.

In the 4D implementation, because floating-point arithmetic is assumed, numerical errors may cause issues such as detecting multiple intersection points where a single point should theoretically exist, or unstable classification when facet vertices lie extremely close to the cutting hyperplane. To address this issue, the following merge procedure was introduced.

First, the intersection point set between the cutting hyperplane and the facet is computed, and duplicates are removed based on a quantization key. Next, among the facet vertices, points that are extremely close to the cutting hyperplane are searched. Let the facet vertex be $\mathbf{v}$, the normal of the cutting hyperplane be $\mathbf{n}$, an arbitrary point on the cutting hyperplane be $\mathbf{p_\text{0}}$, and the tolerance be $\varepsilon$. A point that is extremely close to the cutting hyperplane is defined as a point satisfying Eq.~\eqref{onplane}.

\begin{equation}\label{onplane}
\lvert (\mathbf{p_\text{0}} - \mathbf{v}) \cdot \mathbf{n} \rvert \leq \varepsilon
\end{equation}

If there exists any vertex satisfying Eq.~\eqref{onplane}, the set consisting of the intersection point set and those vertices is treated as the intersection candidate set, and duplicates are removed again based on the quantization key. When an intersection point and a facet vertex are classified under the same quantization key, the intersection coordinates are replaced with the facet vertex coordinates. Although this operation may introduce minor deformation of the mesh, it is positioned as a process to suppress gaps and overlaps in the subdivided mesh.

\subsubsection{Inside-Outside Testing and Result Construction}
\begin{figure*}[t]
    \centering
    \includegraphics[width=\textwidth]{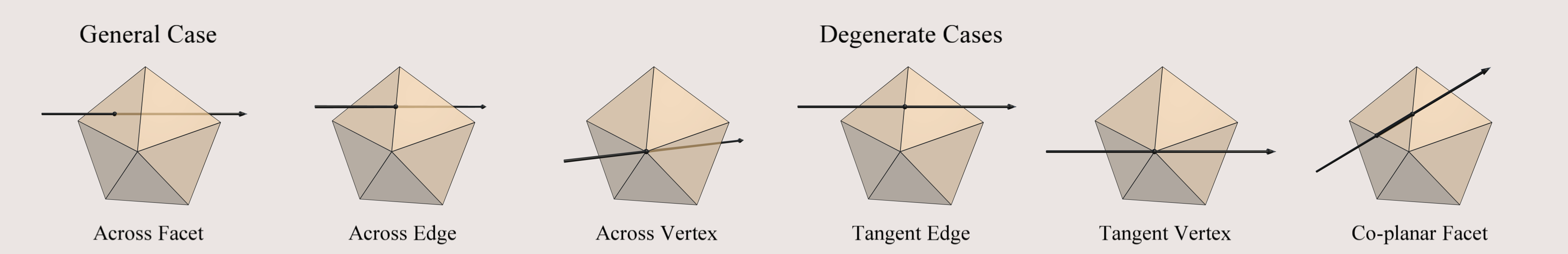}
    \caption{Classification of general and degenerate intersection cases. This figure categorizes the general case, where a ray crosses a single facet, and representative degenerate cases that require precise intersection tests, where the ray passes exactly through a facet, edge, or vertex.}
    \label{fig:booleanCrossAndTangent}
\end{figure*}

To construct the final result of a Boolean operation (union, intersection, or difference) from a set of tessellated facets, it is necessary to classify whether each facet lies inside or outside the other object. For inside-outside classification, this architecture adopts a ray-casting method, which is used in Boolean operations that emphasize interactivity \cite{RobustMeshBooleans}. This method determines whether a point lies inside or outside an object based on the parity of the number of intersections between a ray cast from the point and the surface of the target object.

\begin{figure}[ht]
    \centering
    \includegraphics[width=\columnwidth]{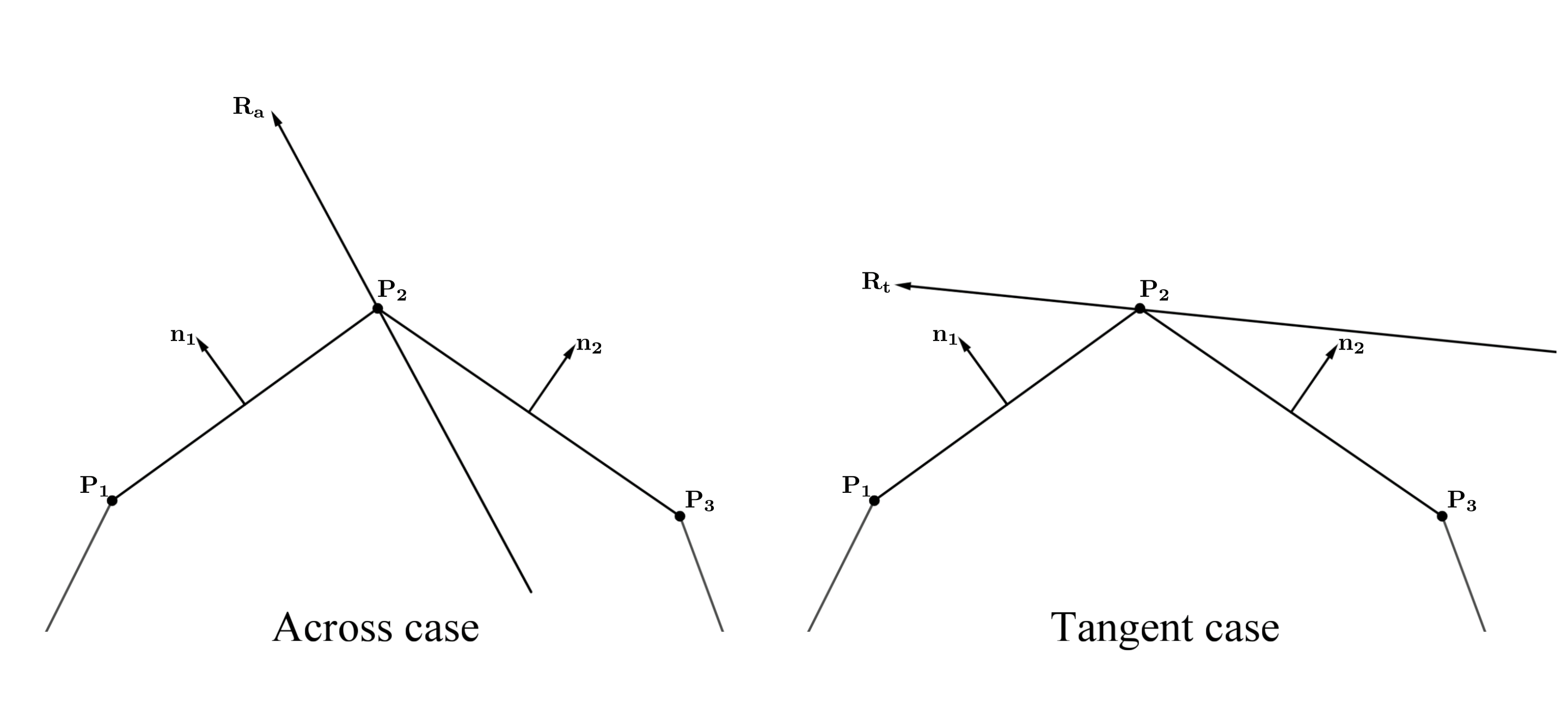}
    \caption{An example of a degenerate case in intersection testing. In the figure, $\mathbf{P_1}$, $\mathbf{P_2}$, and $\mathbf{P_3}$ are the vertices of a 2D polygon; $\mathbf{n_1}$ and $\mathbf{n_2}$ are the outward-facing normal vectors of the edges connected to vertex $\mathbf{P_2}$; and $\mathbf{R_t}$ and $\mathbf{R_a}$ are two different rays passing through $\mathbf{P_2}$. The left diagram shows a piercing case, and the right diagram shows a tangent case.}
    \label{fig:BooleanEdgeCase}
\end{figure}

In the 4D implementation, for the sake of implementation simplicity, the centroid of the facet under consideration is used as the ray origin, and the ray is cast parallel to the $x$-axis. The ray endpoint is set to a point slightly beyond the axis-aligned bounding box (AABB) of the opposing mesh. As illustrated in Fig.~\ref{fig:BooleanEdgeCase}, in degenerate cases where the ray passes through a facet vertex or edge, floating-point arithmetic may cause intersections that should theoretically be a single event at a vertex or edge to be detected as multiple, slightly different intersection points associated with multiple facets sharing that vertex or edge. This can lead to misclassification in inside-outside tests. To address this issue, we quantize the detected intersection points using a tolerance and group points with the same quantization key into a single intersection event.
Next, the nature of this grouped intersection event is determined. Specifically, it is necessary to distinguish whether the ray is piercing the surface or merely touching it.

To do this, we evaluate the dot product of the ray's direction vector and the normal vector of every facet included in the event. If the signs of the dot products are biased to one side (either all positive or all negative), the event is considered to be piercing the surface, and the intersection count is incremented by one. If, on the other hand, the signs are distributed between both positive and negative, the ray is determined to be just tangent to the surface, and it is not included in the intersection count. Based on this inside-outside test executed through this series of processes, the tessellated facets are classified according to the type of Boolean operation, and the final mesh is constructed.

However, in cases where the ray lies on the same plane as a facet, such as the “co-planar facet” case illustrated in Fig.~\ref{fig:booleanCrossAndTangent}, the dot product between the ray direction and the facet normal vector is theoretically zero. As a result, misclassification may occur in sign-based decision methods. Therefore, under such conditions, there is no guarantee that intersections can always be classified correctly. To avoid these degenerate geometric configurations, techniques such as ray perturbation can be considered as candidate approaches.

\subsection{Coordinate Transformations for High-Dimensional Exploration}
\subsubsection{Pose Representation and Management with Geometric Algebra Rotors}
In this architecture, the orientation of an $N$-dimensional object is represented using rotors, in accordance with geometric algebra, which serves as the mathematical foundation described in Section~\ref{overallabstract}.

A rotor is a rotation operator defined by a specific plane of rotation and a rotation angle, which allows the composition of rotations and their application to vectors to be handled as an algebraic product, regardless of the dimension. A bivector $\mathbf{B}$ representing the plane of rotation is parallel to the plane spanned by two vectors, $\mathbf{a}$ and $\mathbf{b}$, and can be defined by their wedge product:

\begin{equation}
    \mathbf{B}=\mathbf{a}\wedge \mathbf{b}
\end{equation}

A rotor is an operator that represents a rotation of an arbitrary angle with respect to this plane. The rotor $R$ is defined using the normalized bivector $\mathbf{\hat{B}}$ as follows:

\begin{equation}
    R=\cos \frac{\theta}{2}+\sin \frac{\theta}{2}\mathbf{\hat{B}}
\end{equation}

The rotation of a vector is then expressed using the rotor as shown in the following equation:

\begin{equation}\label{rotation}
    \mathbf{v}'=R\mathbf{v}\tilde{R}
\end{equation}

Here, $\tilde{R}$ is the reverse of the rotor $R$. If $R$ is normalized, $\tilde{R}$ is equivalent to $R^{-1}$.

In the 4D implementation, users do not manipulate rotors directly; instead, rotation amounts are specified as individual parameters, analogous to Euler angles, for each of the six fundamental rotation planes in 4D space $(xy, xz, xw, yz, yw, zw)$. These parameter values are composed in a prescribed order to internally construct a rotor as the final rotation operator, which is then used for rotation computation.

\subsubsection{View Transformation}\label{ViewTransformation}
This section defines the view transformation in $N$-dimensions, which transforms vertices from the world coordinate system to the camera coordinate system based on the camera's state (position and pose).

\begin{figure}[ht]
    \centering
    \includegraphics[width=\columnwidth]{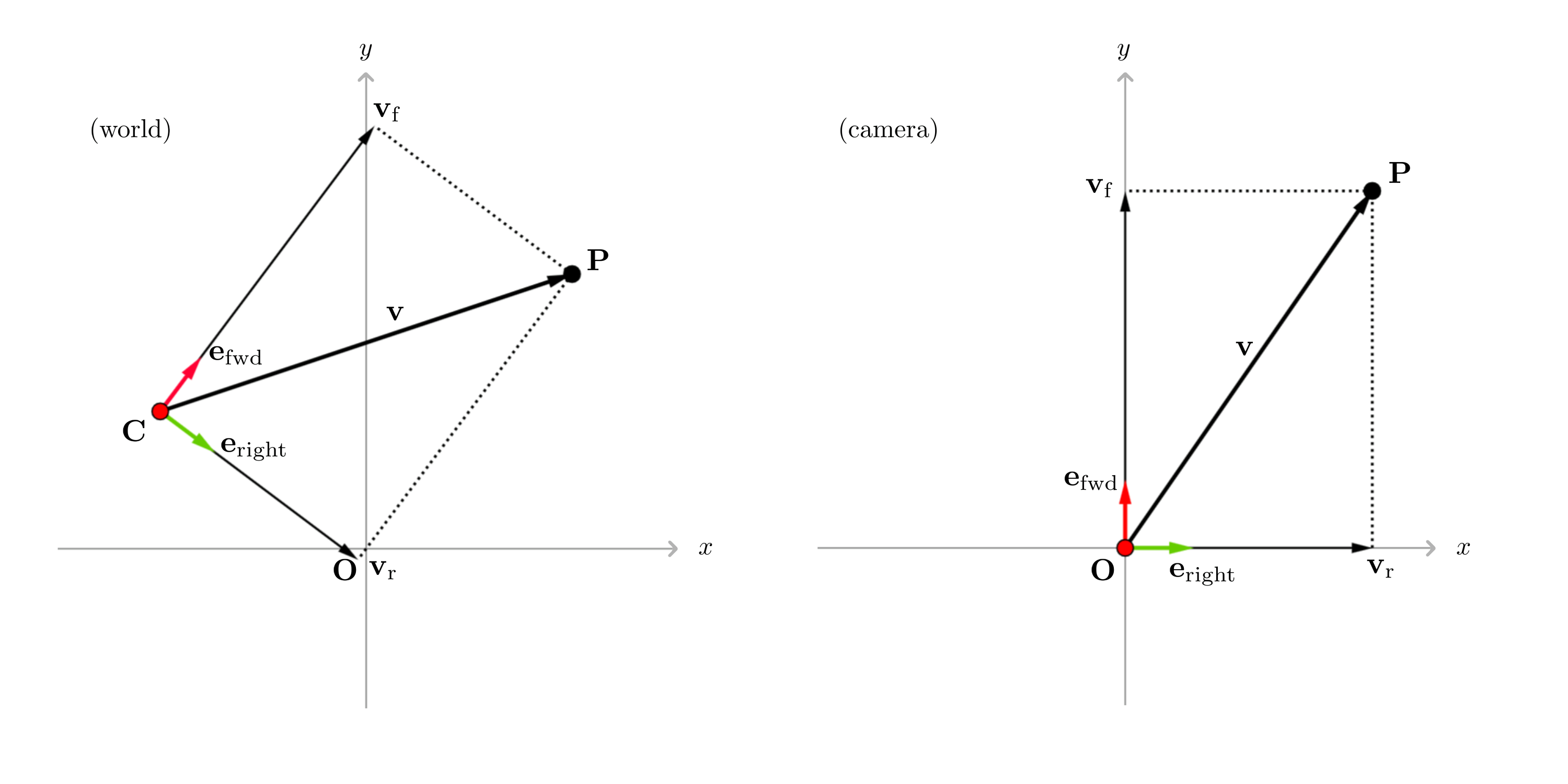}
    \caption{A conceptual diagram of the coordinate transformation from the world coordinate system to the camera coordinate system. Let the camera's position be $\mathbf{C}$ and the target point be $\mathbf{P}$. The relative position vector of the target point, $\mathbf{v}$, is defined as the difference between $\mathbf{P}$ and $\mathbf{C}$. The components obtained by projecting $\mathbf{v}$ onto the basis vectors that represent the camera's orientation, $\mathbf{e}_{\text{right}}$ and $\mathbf{e}_{\text{fwd}}$, directly become the coordinate values of $\mathbf{P}$ in the camera coordinate system. In this setup, the camera's position $\mathbf{C}$ corresponds to the origin of the camera coordinate system.}
    \label{fig:ConvertSight}
\end{figure}

To intuitively understand the geometric essence of this transformation, we consider the 2D space shown in Figure~\ref{fig:ConvertSight} as an example. We denote the position of a target vertex $\mathbf{P}$ in the world coordinate system as $\mathbf{P}^{(\mathrm{W})}$ and the position of the camera $\mathbf{C}$ as $\mathbf{C}^{(\mathrm{W})}$. The coordinate transformation takes $\mathbf{P}^{(\mathrm{W})}$, $\mathbf{C}^{(\mathrm{W})}$, and the rotor $R$ representing the pose as input. In the transformation process, the rotor $R$ representing the camera orientation and the basis vectors of the world coordinate system are first used to rotate the basis vectors according to Eq.~\eqref{rotation}, thereby deriving the basis vectors $\mathbf{e}_{\text{right}}^{(\mathrm{W})}$ and $\mathbf{e}_{\text{fwd}}^{(\mathrm{W})}$ that span the camera coordinate system. Next, the relative position vector $\mathbf{v}^{(\mathrm{W})}$ from the camera's position to the object's vertex is calculated as shown in Equation~\eqref{relvec}.

\begin{equation}\label{relvec}
\mathbf{v}^{(\mathrm{W})} = \mathbf{P}^{(\mathrm{W})} - \mathbf{C}^{(\mathrm{W})}
\end{equation}

Finally, this relative vector $\mathbf{v}^{(\mathrm{W})}$ is projected onto each basis vector via the dot product, and the resulting projection lengths determine the respective components of $\mathbf{P}^{(\mathrm{C})}$.

\begin{equation}\label{projection}
    \begin{aligned}
    \mathbf{P}^{(\mathrm{C})}_x &= \mathbf{v}^{(\mathrm{W})} \cdot \mathbf{e}_{\text{right}}^{(\mathrm{W})}\\
    \mathbf{P}^{(\mathrm{C})}_y &= \mathbf{v}^{(\mathrm{W})} \cdot \mathbf{e}_{\text{fwd}}^{(\mathrm{W})}\\
    \end{aligned}
\end{equation}
    
This entire sequence of processes can be interpreted as two separate geometric operations. Equation~\eqref{relvec} corresponds to a translation that aligns the origin of the world coordinate system with the camera's position. Equation~\eqref{projection} corresponds to a rotation that aligns the coordinate system with the camera's orientation. By using homogeneous coordinates, these two operations can be combined into a single $(N+1) \times (N+1)$ matrix, analogous to the view matrix commonly used in computer graphics. The dot-product-based description in this paper is intended to make the geometric meaning of the rotation operation explicit.

\subsection{Visualization of N-Dimensional Meshes}
\subsubsection{Cross-Sectioning as the Basic Strategy}\label{hyperSlice}
This component generates and renders 3D cross-sections from $N$-dimensional mesh data. The basic strategy for realizing this process is based on the idea of sequentially reducing dimensionality. This is a process of recursively clipping an $N$-dimensional object with an $(N-1)$-dimensional hyperplane, and then clipping the resulting cross-section with an $(N-2)$-dimensional hyperplane. In this paper, we refer to this method of sequentially reducing dimensions as Hierarchical Cross-Sectioning. With this approach, complex cross-section computation problems in arbitrary dimensions are always reduced to a simpler problem: the intersection between an $N$-dimensional facet and an $(N-1)$-dimensional hyperplane.

However, because the information obtained through sequential hyperplane slicing is limited to cross-sections, it involves information loss, and the original shape cannot be uniquely reconstructed from the cross-sections alone. A quantitative evaluation of such perceptual limitations is beyond the scope of this paper.

In the following sections, as a concrete application of this Hierarchical Cross-Sectioning for $N=4$, we will detail the specific algorithm for generating 3D cross-sections from 4D objects and discuss its implementation challenges.

\subsubsection{Facet Clipping Algorithm in the 4D Implementation}\label{facetCutAlgorithm4D}
To generate 3D cross-sections of objects embedded in 4D space, the present implementation computes the geometric intersections between each facet of the 4D mesh, represented as tetrahedra, and an axis-aligned hyperplane $w = w_{\text{slice}}$, where $w_{\text{slice}}$ is an arbitrary constant. This computation is performed for all facets of the input mesh. Cross-section generation based on this type of slicing plane has also been adopted in existing studies on interactive 4D visualization \cite{GL4D}.
Since rotating the object’s orientation via view transformation enables cross-sectional observation from arbitrary directions, this implementation does not employ a procedure to arbitrarily specify the orientation of the cutting hyperplane. Cross-sections are always generated using the axis-aligned plane $w = w_{\text{slice}}$.

First, we determine on which side of the clipping hyperplane each of the facet's four vertices is located. If all vertices are on the same side, the facet does not intersect the hyperplane, and we skip the intersection computation. If the vertices straddle the hyperplane, the intersection point of each constituent edge of the facet with the hyperplane is calculated via linear interpolation. Let the two vertices defining an edge be

\begin{equation}
\mathbf{p_0} = ({x_0}, {y_0}, {z_0}, {w_0})\quad
\mathbf{p_1} = ({x_1}, {y_1}, {z_1}, {w_1})
\end{equation}

The interpolation parameter $t$ for determining the position of the intersection point is then calculated from the $w$-coordinates of both vertices and the $w$-coordinate of the clipping plane, $w_{\text{slice}}$, as shown in Equation~\eqref{eq:intersection_t}.

\begin{equation} \label{eq:intersection_t}
t = \frac{w_{\text{slice}} - w_0}{w_1 - w_0}
\end{equation}
Next, using this parameter $t$, the coordinates of the intersection point $\mathbf{p}_{\text{isect}}$ in 4D space are calculated according to Equation~\eqref{eq:intersection_point}.

\begin{equation} \label{eq:intersection_point}
\mathbf{p}_{\text{isect}} = \mathbf{p}_{0} + t(\mathbf{p}_{1}-\mathbf{p}_{0})
\end{equation}

The 3D coordinates obtained by extracting the $x,y,z$ components of the resulting 4D coordinate $\mathbf{p}_{\text{isect}}$ are directly used as the vertex coordinates constituting the cross-section.

By performing this operation for all edges of facets that intersect the hyperplane, the vertex set of the cross-section is obtained, and polygons are constructed by connecting these vertices. The shape of the cross-section generated by the intersection of a single 4D facet and a 3D hyperplane is, within 3D space, either a triangle with three vertices or a convex polygon with four vertices. For a convex polygon with four vertices, it is divided into two triangles to obtain a set of triangular polygons for rendering.

\subsubsection{Simplicial Decomposition of the Cross-Section}\label{devidecuttedplane}
\begin{figure}[ht]
    \centering
    \includegraphics[width=\columnwidth]{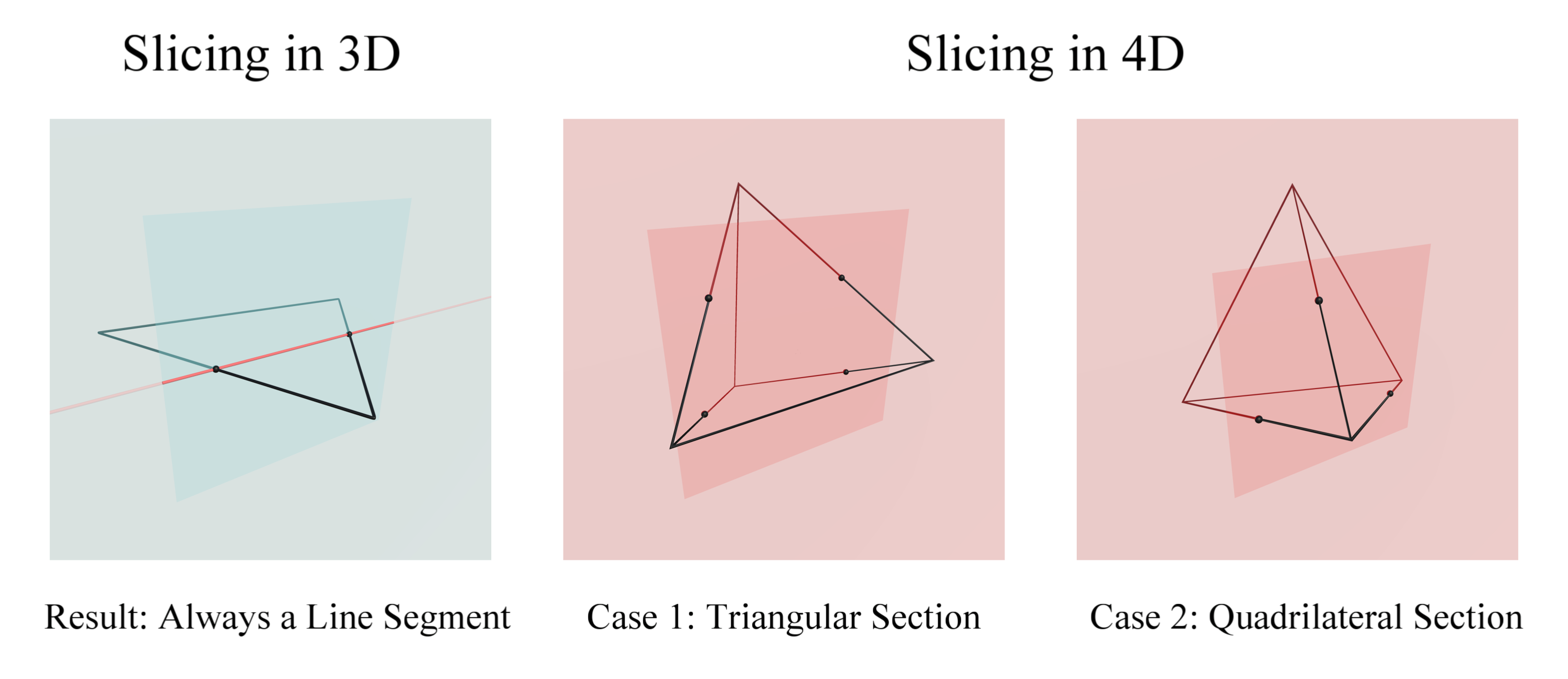}
    \caption{The dimensional dependence of the number of cross-section vertices generated by hyperplane cutting is as follows. In three dimensions, facet cutting produces two vertices, whereas in four dimensions the number of vertices becomes three or four depending on the hyperplane’s relative position. Regarding the color coding in this figure, the blue plane represents the clipping hyperplane, and the red area indicates the portion of it that exists on the dimension being clipped. The darker shades are highlights for visual clarity.}
    \label{fig:Slicing3D4D}
\end{figure}

The cross-sections generated by the aforementioned cutting algorithm form an $(N-2)$-dimensional convex polytope in $N$ dimensions. For subsequent processing, this must be decomposed into a set of $(N-2)$-dimensional simplices. The maximum number of vertices of an $(N-2)$-dimensional cross-section obtained by cutting an $N$-dimensional facet (with $N$ vertices) by a hyperplane is given by ~\eqref{eq:max_vertices}. 

\begin{equation}
    \left\lfloor \frac{N}{2} \right\rfloor \cdot \left\lceil \frac{N}{2} \right\rceil
    \label{eq:max_vertices}
\end{equation}
For example, in 5D, the cross-section of a facet can be a 3D convex polytope with a maximum of $\lfloor5/2\rfloor \cdot \lceil5/2\rceil = 6$ vertices. For simplicial decomposition of such polytopes, the simplex-filling function of Direct Quickhull described in Section~\ref{simplex_filling} is used. By providing the vertex set constituting the cross-section as input to Direct Quickhull, the cross-section is subdivided into a set of $(N-2)$-dimensional simplices.

In the 4D implementation, as illustrated in Fig.~\ref{fig:Slicing3D4D}, the cross-section is a two-dimensional convex polygon with at most four vertices. Therefore, the vertex order is arranged by angle sorting using the cross product, and a four-vertex two-dimensional convex polygon is subdivided into two triangles.

\subsubsection{Determination of Cross-Section Normals and Vertex Winding Order}
In the 4D implementation, when generating 3D cross-sections to be rendered from a 4D object, it is necessary to determine the vertex winding order of the 3D facets. The cutting algorithm described in the previous section produces an unordered set of vertices. If this unordered vertex set is used as is, many rendering pipelines determine front-facing and back-facing polygons based on the input order, which can result in inconsistent face orientations and, consequently, incorrect rendering of the shape.

To address this issue, the present implementation uses the normal vector of the original 4D facet as a reference for the correct orientation. First, the $x,y,z$ components of the 4D facet normal vector are extracted to project the normal vector into 3D space, and this projected vector is defined as the reference normal vector for determining the winding order of the 3D facet. Next, the dot product between this reference normal vector and a provisional normal vector computed by the cross product from the vertex set of the generated 3D facet is evaluated. If the sign of the dot product is negative, the vertex winding order is reversed, thereby ensuring consistent winding of the triangles constituting the cross-section mesh.

\subsection{Interaction Scheme for Interactive Exploration}
\subsubsection{Interaction Scheme Design for the 4D Implementation}
In the 4D implementation, among the six rotational degrees of freedom, the operations handled by the mouse and keyboard are assigned according to the following policy. First, to keep the mapping between inputs and actions as orthogonal as possible, we adopt FPS navigation as the basis, separating translation to the keyboard (WASD) and viewpoint rotation to the mouse. Navigation centered on continuous first-person movement has been evaluated in comparative studies of interaction techniques for exploration tasks \cite{tan2001exploring,mcclymont2011comparison}.

Second, to handle high-degree-of-freedom rotations with limited 2D input, the assignment of horizontal mouse movement ($x$-axis) is switched using a modifier key (LeftControl). In the default state, it is assigned to yaw rotation (the $xz$ plane), and while the modifier key is held down, it is switched to rotation in the $xw$ plane. Such a “spring-loaded mode” (quasimode), which is active only while the key is held, has been discussed as a design that can reduce mode errors compared to persistent modes \cite{hinckley2006springboard}.

\subsubsection{Application of Interaction Design Policies to N Dimensions}
In this section, we describe an interaction scheme obtained by applying the assignment policy used in the aforementioned 4D implementation to $N$ dimensions. Since the rotational degrees of freedom in an $N$-dimensional space increase to $N(N-1)/2$, translation and rotation within the observed 3D cross-section are designed to use dimension-independent operations (WASD/mouse).
In contrast, for rotations outside the observed space, following the same principle as in the 4D implementation, additional degrees of freedom are introduced incrementally while preserving input orthogonality. Specifically, horizontal mouse movement is assigned to rotations outside the observed space, higher-dimensional axes $(w,u,\ldots)$ are selected via combinations of modifier keys, and the corresponding rotation planes $(xw,xu,\ldots)$ are designated as the manipulation targets. This type of switching is consistent with a design policy in which constraints are used to control the degrees of freedom handled by the user \cite{stuerzlinger2010value}.

\subsection{Design Based on Topological and Geometric Separation}\label{TopologyandGeometry}
In physical simulations and related applications, when handling geometry that changes over time while sharing the same topology, a unique vertex array constituting the mesh is typically assumed as input. In contrast, visualization and rendering pipelines treat facets as the minimal unit; therefore, when a collection of facets is represented using vertex sets, duplicate indices arise, and such pipelines assume vertex arrays that include these duplicates as input. As a result, the data formats required by update systems, which demand unique vertex arrays, and by visualization and rendering systems, which assume vertex arrays containing duplicated indices, do not coincide. Under the assumptions of the update and visualization/rendering systems considered in this study, it is therefore not possible to integrate both within a single architecture without separating topology and geometry.

To address this issue, the proposed architecture adopts a design in which, at the initialization stage, a correspondence table (index mapping) between a unique vertex array and a vertex array containing duplicated indices is maintained, derived from the mesh data.

This design allows update systems that take a unique vertex array as input and visualization/rendering systems that assume vertex arrays with duplicated indices to be handled simultaneously within the same architecture. Update processing is completed solely on the unique vertex array, while reflection to visualization and rendering is separated as a transfer process based on the correspondence table.

In the 4D implementation, based on this design, a non-rigid body simulation based on XPBD is integrated, and it is confirmed that mesh vertex coordinates are updated over time and reflected in cross-section generation and rendering. This serves as an implementation example demonstrating that new functionality can be integrated in the four-dimensional implementation without modifying the existing logic. Here, the role of XPBD is limited to serving as an example of a data source that provides dynamically changing geometry, and the validity or accuracy of the physical model is not within the scope of this study.

\section{Plex: An N-Dimensional Mesh Data Exchange Format}
\begin{table*}[t!]
    \centering
    \footnotesize
    \caption{Overview of standard chunks in the \texttt{.plex} format} \label{tab:plex_chunks_overview}
    \begin{tabular}{lcp{10cm}}
        \toprule
        \textbf{Chunk Type} & \textbf{Requirement} & \textbf{Description} \\
        \midrule
        \texttt{'META'} & Required  & Stores metadata for the entire mesh, such as the number of dimensions and vertices. \\
        \texttt{'VERT'}/\texttt{'VERD'} & Required  & Stores the geometry data of vertex coordinates in single or double precision. \\
        \texttt{'FACE'} & Required  & Stores the vertex indices (topology) that constitute the facets. \\
        \texttt{'NORM'}/\texttt{'NRMD'} & Required  & Stores the normal vectors that define the orientation of the facets in single or double precision. \\
        \texttt{'CENT'}/\texttt{'CNTD'} & Optional  & Stores the cached centroid coordinates of facets to reduce computational load. \\
        \bottomrule
    \end{tabular}
\end{table*}

\begin{table*}[t!]
    \centering
    \footnotesize
    \caption{Internal data layout of the \texttt{'META'} chunk}
    \label{tab:meta_chunk_layout}
    \begin{tabular}{clcp{10cm}}
        \toprule
        \textbf{Offset} & \textbf{Size} & \textbf{Type} & \textbf{Description} \\
        \midrule
        \texttt{0x00} & 4 bytes & \texttt{int32} & \textbf{Dimension:} \textit{N} \\
        \texttt{0x04} & 4 bytes & \texttt{uint32} & \textbf{Reserved:} Padded with 0. \\
        \texttt{0x08} & 8 bytes & \texttt{uint64} & \textbf{Vertex Count:} \textit{V} \\
        \texttt{0x10} & 8 bytes & \texttt{uint64} & \textbf{Creation Time:} Unix Time Stamp \\
        \texttt{0x18} & 8 bytes & \texttt{uint64} & \textbf{Modified Time:} Unix Time Stamp \\
        \texttt{0x20} & 1 byte & \texttt{uint8} & \textbf{Precision Flag:} 0=single precision, 1=double precision \\
        \texttt{0x21} & 3 bytes & \texttt{byte[3]} & \textbf{Reserved:} Padded with 0. \\
        \texttt{0x24} & 4 bytes & \texttt{uint32} & \textbf{LSoftware:} Byte length of the following software name. \\
        \texttt{0x28} & Variable & \texttt{char[]} & \textbf{Software Name:} UTF-8 encoded string. \\
        \bottomrule
    \end{tabular}
\end{table*}

\subsection{Background and Challenges}
As described above, when exchanging high-dimensional mesh data across projects, the availability of common formats is limited. This situation can lead to additional effort when sharing data across different implementations or research groups, including data conversion, validation, and interpretation of specifications.

Since the architecture proposed in this study is intended for use by researchers and developers, we define a chunk-based file format, “Plex” (\texttt{.plex}), to enable the exchange of $N$-dimensional mesh data.

\subsection{Relation to Existing Formats and Positioning of Plex}
HDF5 and netCDF are formats designed to store multidimensional array data in a self-describing manner and can handle $N$-dimensional datasets \cite{folk2011overview,rew1990netcdf}. However, because their primary purpose is array storage, they do not directly specify the topology and geometry of $N$-dimensional meshes as part of their specifications.

The VTK file formats are organized as a family of formats for describing datasets that include points, cell elements, and their associated attributes \cite{VTKFileFormats}. XDMF adopts a model for representing data such as meshes based on XML-based metadata descriptions and references to external data \cite{XDMFModelFormat}. In contrast, this study focuses on defining, as a single specification, the minimal elements commonly handled by mesh data parameterized by the dimension $N$. These elements are topology and geometry. This focus does not align with the design objectives of these formats.

Based on the above considerations, this study does not aim to replace existing formats; instead, Plex is positioned as a format that explicitly describes the minimal elements (topology and geometry) commonly represented in $N$-dimensional mesh data.

\subsection{Dual-Format Strategy}
In research workflows that handle $N$-dimensional data, two different requirements may coexist: inspection and validation of data contents at early stages, and processing performance for large-scale computations. To address these requirements, Plex adopts a design that combines a human-readable JSON format and a chunk-based binary format \texttt{.plex}. The human-readable format is used for content inspection and debugging, while the binary format is used to improve efficiency in loading and saving.

The present implementation includes a one-way converter from \texttt{.plex} to the JSON format. This enables a workflow in which data are normally stored in the binary format and exported to the human-readable format only when necessary. In this paper, this conversion is presented as a minimal means for content inspection and third-party verification of Plex data, while providing converters to VTK-family formats or HDF5 is considered out of scope and positioned as future work.

\subsection{.plex Binary Format Specification}
The \texttt{.plex} format is defined as a chunk-based format consisting of a sequence of self-describing chunks. A file comprises a global header and at least four self-describing chunk sequences shown in Table~\ref{tab:plex_chunks_overview}, and the internal structure of the \texttt{META} chunk is shown in Table~\ref{tab:meta_chunk_layout}. Each chunk contains a four-character chunk type, a data length, a data payload, and a CRC32 checksum for verifying data integrity. The data payloads of the primary chunks, except for the \texttt{META} chunk, are defined as follows.

In the \texttt{FACE} chunk, which defines topology, the total number of facets is recorded as a 64-bit unsigned integer, followed by the vertex index lists (arrays of 32-bit unsigned integers) that constitute each facet. The \texttt{VERT} or \texttt{VERD} chunks that store vertex coordinates contain no metadata such as counts; for an $N$-dimensional mesh with $V$ vertices, they consist solely of a contiguous array of $V \times N$ numerical values. The data types are 32-bit single precision and 64-bit double precision, respectively. The data layouts of the \texttt{NORM}/\texttt{NRMD} chunks and optional \texttt{CENT}/\texttt{CNTD} chunks likewise conform to the formats of the \texttt{VERT}/\texttt{VERD} chunks.

From a compatibility perspective, parsers compliant with the specification are required to be able to skip unknown chunk types, allowing loading to continue as long as the existing required chunks can be interpreted. In addition, assuming future extensions such as additional attributes or auxiliary information to be introduced as chunks, a naming convention is defined in which identifiers containing at least one lowercase letter are reserved for custom chunks. Here, “compatibility” refers to forward compatibility within the Plex specification, whereby loading can proceed using only known chunks while ignoring unknown ones, and does not imply binary compatibility with VTK-family formats or HDF5.

\begin{figure*}[t]
    \centering
    \includegraphics[width=\textwidth]{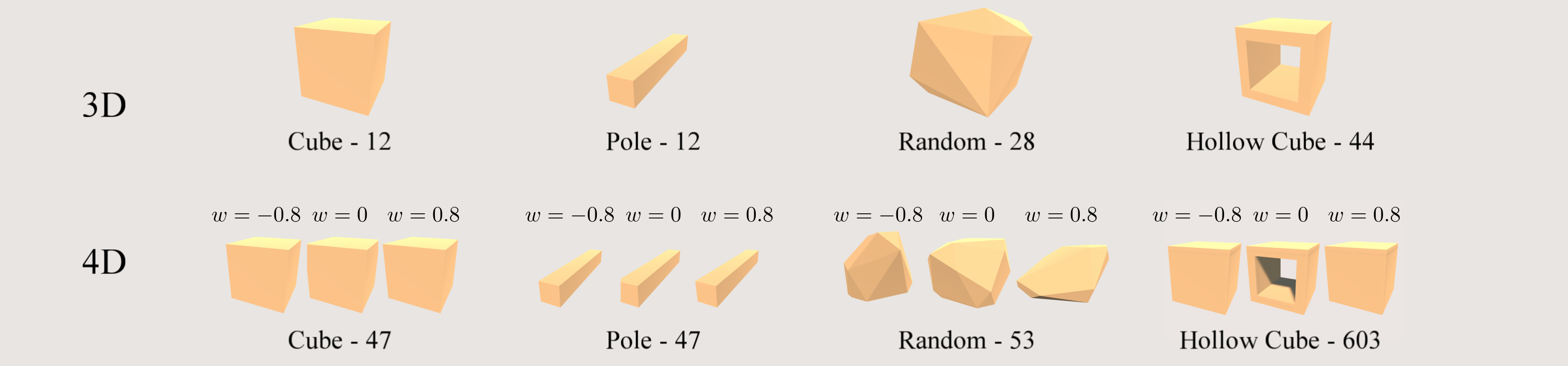}
    \caption{The set of base objects used in the experiments. The top row shows the 3D objects, and the bottom row shows the 4D objects. The numerical value shown to the right of each name indicates the mesh size (number of facets). 4D objects are displayed as 3D cross-sections obtained using the hyperplane slicing function described in this paper. As an example, cross-sections at three $w$ coordinates $(w=-0.8,0,0.8)$ are arranged from left to right.}
    \label{fig:experimentObject}
\end{figure*}

\section{Experiments and Evaluation}\label{experiment}
\subsection{Experimental Setup}
The performance evaluation in this chapter, for the 4D implementation, does not aim to compare execution speed with existing offline methods. Instead, it aims to evaluate the scalability of performance with respect to the size of the target meshes (e.g., the number of facets) within the proposed CPU-based architecture. All implementation and performance evaluations in this study were conducted in the following environment.

\begin{itemize}
    \setlength{\itemsep}{0pt}
    \setlength{\parskip}{0pt}
    
    \item[] \hspace{-1.5em} \textbf{Hardware Environment:}
        \begin{itemize}[leftmargin=0.5em]
            \item \textbf{CPU:} Intel Core i7-12700H (2.30 GHz, 14 Cores)
            \item \textbf{Memory:} \texttt{16 \si{\giga\byte} RAM}
            \item \textbf{GPU:} Intel Iris Xe Graphics (Integrated, Shared Memory)
        \end{itemize}
    
    \item[] \hspace{-1.5em} \textbf{Software Environment:}
        \begin{itemize}[leftmargin=0.5em]
            \item \textbf{OS:} Windows 11 Home (64-bit)
            \item \textbf{Development Environment:} Unity 2022.3 LTS (2022.3.62f2)
            \item \textbf{Programming Language:} C\#
        \end{itemize}
\end{itemize}

\subsection{Algorithmic Performance Evaluation}
\begin{table}[H]
    \centering
    \footnotesize
    \caption{Performance evaluation of the Direct Quickhull implementation. This table shows the average processing time for convex hull construction with randomly placed vertices (vertex count: 10 to 1000) in 3D and 4D spaces. The processing time is the average of 100 trials.}
    \label{tab:quickhull}
    \begin{tabular}{crd{3.2} d{3.2}} 
        \toprule
        \textbf{Dimension} & \textbf{Vertices} & \multicolumn{1}{c}{\textbf{Time (\si{\milli\second})}} & \multicolumn{1}{c}{\textbf{Avg. Facets}}\\
        \midrule
        3D & 10 & 0.05 & 13.52\\
        3D & 20 & 0.12 & 25.32\\
        3D & 50 & 0.35 & 48.20\\
        3D & 100 & 0.88 & 73.34\\
        3D & 200 & 2.50 & 110.66\\
        3D & 500 & 9.43 & 185.24\\
        3D & 1000 & 26.95 & 268.32\\
        
        \midrule
        4D & 10 & 0.21 & 20.38\\
        4D & 20 & 0.56 & 43.11\\
        4D & 50 & 1.64 & 100.02\\
        4D & 100 & 3.84 & 174.32\\
        4D & 200 & 7.97 & 275.43\\
        4D & 500 & 27.43 & 496.94\\
        4D & 1000 & 75.99 & 779.91\\
        
        \bottomrule
    \end{tabular}
\end{table}

\begin{table*}[t]
    \centering
    \footnotesize
    \caption{Performance evaluation results for Boolean operations. For combinations of the basic objects shown in Fig.~\ref{fig:experimentObject}, union, intersection, and difference operations were performed in three- and 4D spaces. Each entry reports the processing time averaged over three trials (\si{\milli\second}) and the size of the generated mesh (number of facets).}
    \label{tab:boolean}
    \rowcolors{2}{myLightGray}{myWhite}
    \begin{tabular}{cllc d{3} d{3}c d{3} d{3}c d{3} d{3}}
        \toprule
        \multicolumn{3}{c}{\textbf{Input}} && \multicolumn{2}{c}{\textbf{Union}} && \multicolumn{2}{c}{\textbf{Intersection}} && \multicolumn{2}{c}{\textbf{Difference}} \\

        \cmidrule(lr){1-3} \cmidrule(lr){5-6} \cmidrule(lr){8-9} \cmidrule(lr){11-12}
        
        {\textbf{Dimension}} & \textbf{ObjA} & \textbf{ObjB} &&
        \multicolumn{1}{c}{\textbf{Time (\si{\milli\second})}} & \multicolumn{1}{c}{\textbf{Size}} &&
        \multicolumn{1}{c}{\textbf{Time (\si{\milli\second})}} & \multicolumn{1}{c}{\textbf{Size}} &&
        \multicolumn{1}{c}{\textbf{Time (\si{\milli\second})}} & \multicolumn{1}{c}{\textbf{Size}}\\
        \midrule

        3D & Cube & Cube && 15.42 & 50 && 13.47 & 26 && 14.47 & 38\\
        3D & Cube & Pole && 11.55 & 56 && 10.62 & 24 && 12.43 & 36\\
        3D & Cube & Random && 20.87 & 87 && 18.60 & 59 && 20.46 & 79\\
        3D & Cube & Hollow Cube && 14.49 & 104 && 11.82 & 28 && 14.65 & 88\\
        3D & Pole & Pole && 13.75 & 64 && 11.46 & 32 && 10.95 & 48\\
        3D & Pole & Random && 25.39 & 86 && 20.78 & 38 && 19.72 & 74\\
        3D & Pole & Hollow Cube && 16.90 & 156 && 12.86 & 84 && 18.29 & 144\\
        3D & Random & Random && 33.65 & 98 && 28.24 & 54 && 24.50 & 78\\
        3D & Random & Hollow Cube && 21.56 & 126 && 20.94 & 56 && 17.82 & 94\\
        3D & Hollow Cube & Hollow Cube && 18.93 & 170 && 14.92 & 66 && 14.97 & 118\\
        \midrule
        4D & Cube & Cube && 323.44 & 1023 && 291.94 & 301 && 333.86 & 694\\
        4D & Cube & Pole && 205.06 & 768 && 186.23 & 299 && 176.74 & 324\\
        4D & Cube & Random&& 2091.09 & 6142 && 2081.24 & 3777 && 2079.20 & 4143\\
        4D & Cube & Hollow Cube && 682.97 & 3017 && 656.85 & 1471 && 661.75 & 2693\\
        4D & Pole & Pole && 139.65 & 857 && 128.59 & 480 && 129.56 & 631\\
        4D & Pole & Random && 1225.35 & 2095 && 1213.80 & 1123 && 1237.35 & 1655\\
        4D & Pole & Hollow Cube && 638.94 & 3357 && 626.98 & 1353 && 639.27 & 3402\\
        4D & Random & Random && 2175.25 & 4449 && 2126.16 & 2326 && 2157.43 & 3114\\
        4D & Random & Hollow Cube && 2406.69 & 9633 && 2332.84 & 5292 && 2418.27 & 9181\\
        4D & Hollow Cube & Hollow Cube && 1598.95 & 7973 && 1559.87 & 4763 && 1576.44 & 6801\\
        \bottomrule
    \end{tabular}
\end{table*}

\begin{table*}[t]
    \centering
    \footnotesize
    \caption{Interactive performance evaluation. Rendering performance is shown for a scene in which 1–20 4D hypercubes, each with 47 facets, are placed, while continuous interaction (translation and viewpoint rotation in the $xz$ and $xw$ planes) is performed for 30 seconds. As performance metrics, the average FPS, minimum FPS, and the 99th percentile frametime (\si{\milli\second}) are reported. All values are averages over three 30-second measurement runs.}
    \label{tab:FPS}
    \begin{tabular}{cccccc} 
        \toprule
        \textbf{Objects (Total Facets)} & \textbf{Average FPS} & \textbf{Minimum FPS} & \textbf{99th Percentile Frametime (\si{\milli\second})}\\
        \midrule
        1(47) & 77.9 & 29.8 & 19.9\\
        5(235) & 58.3 & 27.9 & 23.4\\
        10(470) & 38.6 & 16.8 & 32.5\\
        15(705) & 27.0 & 14.4 & 46.9\\
        20(940) & 21.2 & 13.1 & 58.1\\
        
        \bottomrule
    \end{tabular}
\end{table*}

\subsubsection{Convex Hull Mesh Generation Performance}
The measurement results are shown in Table~\ref{tab:quickhull}. For point sets with up to 50 vertices, which correspond to the intended input of this implementation, the processing time was less than \SI{1.7}{\milli\second} in all tested dimensions within the measured range.

In this implementation, convex hull generation is a processing stage whose computational cost varies depending on the input size within the architecture. To present baseline data for the architectural design regarding how convex hull generation time increases with input size, measurements were also conducted over the range of 100--1000 vertices as a scalability test beyond the intended input scale.

\subsubsection{Boolean Operation Performance}
The measurement results are shown in Table~\ref{tab:boolean}. In 3D space, for the combinations tested, the computation time was generally on the order of several tens of milliseconds. In contrast, in 4D space the computation time increased, reaching approximately \SI{300}{\milli\second} for Cube–Cube cases and exceeding \SI{2}{\second} for Random–Random cases in some instances. Numerical-error-induced holes or cracks in the mesh were not observed by visual inspection.

\subsection{Rendering Performance}
The measurement results are shown in Table~\ref{tab:FPS}. Within the measured range, the FPS decreased as the total number of facets increased. Under the experimental conditions used in this study, when the total number of facets was 470 (corresponding to 10 objects), the average FPS was approximately 38.6. These values are recorded as reference values for the present implementation and environment.

\subsection{Practical Scope of Application and Usage Scenarios}

In this section, based on the measurement results of the 4D implementation, we describe the types of applications that can be realistically supported by the current implementation.

Regarding visualization, in the present experiments, the frame rate decreased as the number of mesh facets increased. This may result in a loss of usability for continuous interaction. Therefore, when using this implementation for exploration or observation with continuous operations, a practical guideline is to keep the total number of facets at approximately 470 or fewer.

For Boolean operations, immediate per-frame updates are not required, and a workflow of “operation → waiting for a certain period → result confirmation → next operation” can be viable. In the experiments, processing time tended to increase as complexity grew, and when it reached the order of several seconds, the pace of trial-and-error interaction was degraded. For this reason, the present implementation is realistically suited to research use cases or prototyping scenarios involving a small number of edits or validations, where waiting for results is acceptable.
In contrast, usage scenarios that require continuous immediate responses, such as editing tasks that repeatedly apply Boolean operations at high frequency to large meshes with many facets, may become impractical with this implementation due to increased processing time as the number of facets grows. Accordingly, such scenarios are positioned outside the scope of the implementation and evaluation presented in this paper.

However, as a workflow-level mitigation strategy, a proxy-based approach can be considered as one possible option. In this approach, trial-and-error operations are performed on simplified meshes, and the operations are applied to high-resolution models only at the final stage.

\subsection{Discussion and Limitations}
The definition of interactivity adopted in this paper holds regardless of computation time; however, practical usability is constrained as computation time increases. The experimental results suggest that when handling meshes with a large number of facets, processing time can reach the order of several seconds, leading to the existence of a practical transition region in which usage patterns that proceed by confirming results after each operation become constrained by waiting time.

It should be emphasized that this transition region does not represent a strict threshold, but rather a practical consideration based on the experimental conditions of this study. Accordingly, this paper does not generalize acceptable response times or practical boundaries as a single numerical value.

The present implementation employs approximate methods based on floating-point arithmetic, and intersection tests as well as inside-outside classification may become unstable depending on degenerate configurations and the choice of tolerances. Therefore, this paper explicitly states the limits of reliability in terms of these dependencies, and does not specify failure conditions as a single numerical boundary.

The experiments in this implementation focus on computational performance and its scalability, and do not include qualitative evaluations such as user studies or structured feedback from domain researchers.

In addition, the visualization-related limitations, namely information loss and dependence on sampling, are described in Section~\ref{hyperSlice}; quantitative evaluation of these limitations is beyond the scope of this paper.

\section{Conclusion}
In this paper, we proposed a unified architecture for visualization and simulation based on a design targeting $N$-dimensional space. We presented a configuration that integrates multiple processes, including convex hull mesh generation, Boolean operations, coordinate transformations, and hyperplane slicing, into a single software architecture, and adopted the separation of topology and geometry as a design principle. In addition, an interaction scheme for high-dimensional exploration was designed. Furthermore, as one example of data representation for $N$-dimensional meshes, this study designed the Plex data format and defined its specification.

In the 4D implementation, a non-rigid body simulation based on XPBD was implemented within the same architecture, and it was confirmed that updates to mesh vertex coordinates are reflected in cross-section generation and rendering.

\section{Future Work}
Although the algorithms used in this study are described in a manner independent of the dimensionality $N$, performance evaluation and implementation-level verification are limited to four dimensions.

Future work includes examining the impact of increased computational cost due to the growth of combinatorial complexity in intersection tests and simplicial subdivision for algorithms in five or more dimensions; ensuring the topological correctness of meshes generated by Direct Quickhull; investigating more numerically robust and physically accurate $N$-dimensional non-rigid physical models; exploring integration using other phenomena, such as fluid simulations, as input examples; and considering GPU implementations for components that are suitable for parallelization among the processes implemented in a CPU-based manner in this paper. In addition, conducting comparative evaluations with existing computational geometry libraries and previously reported methods, and quantitatively characterizing the trade-offs among computation time, memory usage, and robustness to degenerate configurations, also remain as future tasks.
Furthermore, conducting qualitative evaluations also remains as future work. Such evaluations include collecting structured feedback from users and domain researchers on representative exploration and editing tasks.

The design and the 4D implementation presented in this study may also serve as candidates for educational and entertainment applications aimed at understanding and manipulating high-dimensional phenomena.

\section*{Acknowledgments}
We would like to express our sincere gratitude to Mr. Hiroaki Naito of Waseda Osaka High School for his guidance throughout this research. We also wish to express our deepest appreciation to Professor Yasushi Homma of Waseda University for his valuable professional advice on the mathematical expressions related to geometric Algebra used in this paper. Finally, we would like to thank the anonymous reviewers for their constructive comments, which helped to improve this manuscript.

\bibliographystyle{plain}
\bibliography{references} 

\end{document}